\RequirePackage{fix-cm}
\documentclass[smallextended]{svjour3}       
\smartqed  
\usepackage{graphicx}
\usepackage{latexsym,amssymb,amsmath,color,bm,url,natbib}

\newcommand{\p}{{\rm Pr}}

\begin{document}

\title{Smoothness-constrained model for nonparametric item response theory
}

\titlerunning{Smoothness-constrained model for nonparametric IRT}        

\author{Toshiki Sato \and Yuichi Takano
}


\institute{
Toshiki Sato \at
Graduate School of Systems and Information Engineering, University of Tsukuba, 1-1-1 Tennodai, Tsukuba-shi, Ibaraki 305-8573, Japan
\and
Yuichi Takano \at
School of Network and Information, Senshu University, 2-1-1 Higashimita, Tama-ku, Kawasaki-shi, Kanagawa 214-8580, Japan\\
}

\date{Received: date / Accepted: date}

\maketitle

\begin{abstract} 
This paper is concerned with the nonparametric item response theory (NIRT) for estimating item characteristic curves (ICCs) and latent abilities of examinees on educational and psychological tests. In contrast to parametric models, NIRT models can estimate various forms of ICCs under mild shape restrictions, such as the constraints of monotone homogeneity and double monotonicity. However, NIRT models frequently suffer from estimation instability because of the great flexibility of nonparametric ICCs, especially when there is only a small amount of item-response data. To improve the estimation accuracy, we propose a novel NIRT model constrained by monotone homogeneity and smoothness based on ordered latent classes. Our smoothness constraints avoid overfitting of nonparametric ICCs by keeping them close to logistic curves. We also implement a tailored expectation--maximization algorithm to calibrate our smoothness-constrained NIRT model efficiently. We conducted computational experiments to assess the effectiveness of our smoothness-constrained model in comparison with the common two-parameter logistic model and the monotone-homogeneity model. The computational results demonstrate that our model obtained more accurate estimation results than did the two-parameter logistic model when the latent abilities of examinees for some test items followed bimodal distributions. Moreover, our model outperformed the monotone-homogeneity model because of the effect of the smoothness constraints. 

\end{abstract}

\section{Introduction}\label{sec:1}
Item response theory (IRT) is a family of statistical measurement methods for educational and psychological tests. In IRT models, the characteristics of each test item are examined based on the item characteristic curve (ICC), which expresses the probability of a correct answer as a function of the latent abilities of the examinees. Indeed, many testing companies use IRT models for the design, analysis, and scoring of tests.

This paper is focused on nonparametric item response theory (NIRT) models~(\citealt{JuSi01,SiMo02,St87,VaHa13}). In contrast to parametric item response theory (PIRT) models in which the ICCs are defined by parametric functions (e.g., logistic curves or normal ogives), NIRT models are capable of estimating various forms of ICCs under mild shape restrictions, such as the monotone-homogeneity constraint~(\citealt{Me65,Mo71}) and the double monotonicity constraint~(\citealt{Mo71,MoLe82}). It has been demonstrated that PIRT models do not always fit the data well~(\citealt{DuCo01,Jo07,Ra91}), under which circumstances NIRT models have a clear advantage. They are also useful for evaluating the data quality~(\citealt{MeTe15}) and the goodness of fit of PIRT models~(\citealt{DuCo01,LiWe15,LiWe14}). 

The existing methods for estimating nonparametric ICCs include regression splines~(\citealt{Jo07,Ra88,RaAb89,RaWi91,RoWa02}), kernel smoothing~(\citealt{Du97,GuSi11,LuRo15,Ra91}), isotonic regression~(\citealt{Le07}), the finite mixture model~(\citealt{MoKa13}), and monotonic polynomial regression~(\citealt{FaCa16,LiBr15}). If the latent abilities of the examinees  are represented as ordered latent classes, NIRT models are categorized as ordered latent class models~(\citealt{Cr90,Cr91,LiVe12,Va02,Ve01}). The expectation--maximization (EM) algorithm~(\citealt{Cr90,Cr91,RoWa02,Ve01}) and Markov chain Monte Carlo (MCMC) method~(\citealt{Jo07,KaSh04,Le09,LiVe12,MoCa14,Va02}) have been employed to estimate both the nonparametric ICCs and the latent abilities of examinees. However, NIRT models frequently suffer from estimation instability because of the great flexibility of nonparametric ICCs, especially if there is only a small amount of item-response data~(\citealt{Mo01}).

Various shape restrictions have been proposed to prevent overfitting with nonparametric regression models~(\citealt{Ra05,Si12}). To improve the estimation accuracy of NIRT models, we make effective use of the smoothness constraints on the nonparametric ICCs. More specifically, we propose an NIRT model with monotone homogeneity and smoothness constraints based on the ordered latent classes. Our smoothness constraints keep each nonparametric ICC close to a logistic curve, and thus offer advantages to both PIRT and NIRT models, namely stability and flexibility. To the best of our knowledge, no existing study has incorporated such smoothness constraints into a monotone-homogeneity NIRT model. In addition, we implement a tailored EM algorithm to calibrate our smoothness-constrained NIRT model efficiently. 

We conducted computational experiments to assess the effectiveness of our smoothness-constrained model in comparison with the common two-parameter logistic model and the monotone-homogeneity model. The computational results demonstrate that our model delivered the best estimation performance in many cases. In other words, the smoothness constraints were very effective in enhancing the estimation accuracy of the NIRT models.

The remainder of this paper is organized as follows. In Sect.~\ref{sec:2}, we present the monotone-homogeneity model for estimating the monotonically increasing ICCs and the ability classes of examinees. In Sect.~\ref{sec:3}, we formulate our smoothness-constrained model and EM algorithm. In Sect.~\ref{sec:4}, we report the computational results, and in Sect.~\ref{sec:5} we conclude the paper with a brief summary of our work. 

\section{Monotone-homogeneity model}\label{sec:2}
In this section, we pose our monotone-homogeneity model by following \cite{TaTs16}. Let us denote by $I$ a set of examinees and by $J$ a set of dichotomously scored question items on a test. The test results are given as the binary item-response data 
\[
\bm{U} := (u_{ij})_{(i,j) \in I \times J} \in \{0,1\}^{|I| \times |J|}, 
\]
where $u_{ij}=1$ if the $i$th examinee provided a correct answer to the $j$th item, or $u_{ij}=0$ otherwise. We make the following assumptions throughout this paper:
\begin{description}
\item[\sf Unidimensionality:] the latent abilities of all examinees are evaluated unidimensionally. 
\item[\sf Local Independence:] item responses are conditionally independent of each other given an individual latent ability. 
\end{description}
Additionally, the latent abilities of examinees are represented as ordered latent classes denoted by $T$. 

The nonparametric ICCs of test items are defined by the decision variable 
\[
\bm{X} := (x_{jt})_{(j,t) \in J \times T}, 
\]
where $x_{jt}$ is the probability of the $j$th item being answered correctly by examinees in the $t$th ability class. These nonparametric ICCs are usually estimated subject to monotone-homogeneity constraints~(\citealt{Me65,Mo71}), which require that the probability of a correct answer increases monotonically with ability class: 
\begin{align}
& x_{jt} \le x_{j,t+1} \quad ((j,t) \in J \times T), \label{con:mh1} \\
& 0 \le x_{jt} \le 1 \quad ((j,t) \in J \times T). \label{con:mh2} 
\end{align}

The ability classes of examinees are represented by the decision variable 
\[
\bm{Y} := (y_{it})_{(i,t) \in I \times T}, 
\]
where $y_{it} = 1$ if the $i$th examinee possesses the latent ability of the $t$th class, or $y_{it} = 0$ otherwise. The following constraints guarantee that only one ability class is assigned to each examinee:
\begin{align}
& \sum_{t \in T} y_{it} = 1 \quad (i \in I), \label{con:asg1} \\
& y_{it} \in \{0,1\} \quad ((i,t) \in I \times T). \label{con:asg2} 
\end{align}

Given $\bm{x}_{j \cdot} := (x_{jt})_{t \in T}$ and $\bm{y}_{i \cdot} := (y_{it})_{t \in T}$, the probability of receiving response $u_{ij} \in \{0,1\}$ is expressed as 
\begin{align} \notag
\p(u_{ij} \mid \bm{x}_{j \cdot},\bm{y}_{i \cdot}) := \prod_{t \in T} \left( (x_{jt})^{u_{ij}} (1 - x_{jt})^{1 - u_{ij}} \right)^{y_{it}}.
\end{align}
From the assumption of local independence, the probability of the $i$th examinee giving the response $\bm{u}_{i \cdot} := (u_{ij})_{j \in J}$ is expressed as 
\begin{align} \notag
\p(\bm{u}_{i \cdot} \mid \bm{X},\bm{y}_{i \cdot}) := \prod_{j \in J} \p(u_{ij} \mid \bm{x}_{j \cdot},\bm{y}_{i \cdot}). 
\end{align}
Because the responses of different examinees are independent, the probability of receiving item response $\bm{U}$ from all the examinees is given by 
\begin{align}
\p(\bm{U} \mid \bm{X},\bm{Y}) 
& := \prod_{i \in I} \p(\bm{u}_{i \cdot} \mid \bm{X},\bm{y}_{i \cdot}) \notag \\
& = \prod_{(i,j,t)  \in I \times J \times T} \left( (x_{jt})^{u_{ij}} (1 - x_{jt})^{1 - u_{ij}} \right)^{y_{it}}. \notag
\end{align}
By treating $\bm{X}$ and $\bm{Y}$ as decision variables, the log-likelihood function is defined as follows: 
\begin{align} 
\ell(\bm{X},\bm{Y} \mid \bm{U}) 
& := \log \p(\bm{U} \mid \bm{X},\bm{Y}) \notag \\
& = \sum_{(i,j,t)  \in I \times J \times T} y_{it} \left( u_{ij} \log x_{jt} + (1 - u_{ij}) \log (1 - x_{jt}) \right). \label{eq:llf}
\end{align}

Consequently, the monotone-homogeneity model estimates $\bm{X}$ and $\bm{Y}$ so that the log-likelihood function~\eqref{eq:llf} is maximized subject to constraints~\eqref{con:mh1}--\eqref{con:asg2}:
\begin{align}
\mathop{\mbox{maximize}}_{\scriptstyle \bm{X},\bm{Y}} & \quad \sum_{(i,j,t)  \in I \times J \times T} y_{it} \left( u_{ij} \log x_{jt} + (1 - u_{ij}) \log (1 - x_{jt}) \right) \label{obj:MHM} \\
\mbox{subject~to} 
& \quad x_{jt} \le x_{j,t+1} \quad ((j,t) \in J \times T), \label{con1:MHM} \\
& \quad 0 \le x_{jt} \le 1 \quad ((j,t) \in J \times T), \label{con2:MHM} \\
& \quad \sum_{t \in T} y_{it} = 1 \quad (i \in I), \label{con3:MHM} \\
& \quad y_{it} \in \{0,1\} \quad ((i,t) \in I \times T). \label{con4:MHM} 
\end{align}

\section{Smoothness-constrained model}\label{sec:3}
In this section, we firstly formulate our smoothness-constrained model and then describe an EM algorithm for model estimation. 

\subsection{Smoothness constraints}
To express our smoothness constraints, we use the logistic function 
\begin{align}
\lambda(w) := \frac{1}{1 + \exp(-w)} 
\label{eq:logf}
\end{align}
and the additional decision variable 
\[
\bm{W} := (w_{jt})_{(j,t) \in J \times T}. 
\]
The ICCs are then defined as $x_{jt} = \lambda(w_{jt})$; that is, $\lambda(w_{jt})$ denotes the probability of the $j$th item being answered correctly by examinees in the $t$th ability class. Because the logistic function increases monotonically from zero to one, the monotone-homogeneity constraints on $\lambda(w_{jt})$ are written as 
\begin{align}
w_{jt} \le w_{j,t+1} \quad ((j,t) \in J \times T). 
\label{con:smh}
\end{align}
The smoothness constraints on the nonparametric ICCs are posed as follows: 
\begin{align}
\sum_{t \in T} |w_{j,t+2} - 2w_{j,t+1} + w_{jt}| \le \gamma \quad (j \in J), \label{con:sc}
\end{align}
where $\gamma \ge 0$ is a user-defined parameter. If $\gamma$ is sufficiently large, constraints~\eqref{con:sc} are invalidated. Conversely, if $\gamma = 0$, constraints~\eqref{con:sc} are equivalent to 
\begin{align}
w_{j,t+2} - w_{j,t+1} = w_{j,t+1} - w_{jt} \quad ((j,t) \in J \times T), 
\end{align}
which imply that for each $j \in J$, $\{w_{jt} \mid t \in T\}$ is a set of equally spaced points. 

\begin{figure}[tbh]
\centering
\begin{tabular}{ccc}
\includegraphics[scale=0.6]{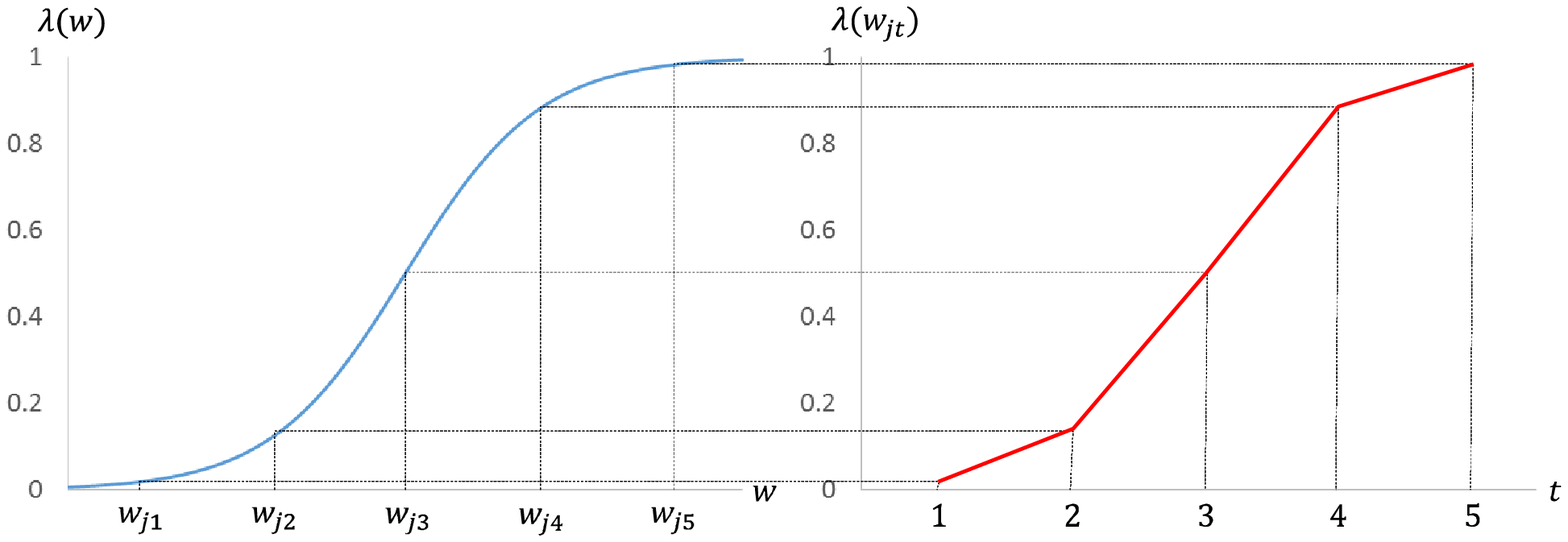} \\ (a) $(w_{j1},w_{j2},w_{j3},w_{j4},w_{j5}) = (-4,-2,0,2,4)$ \\\\
\includegraphics[scale=0.6]{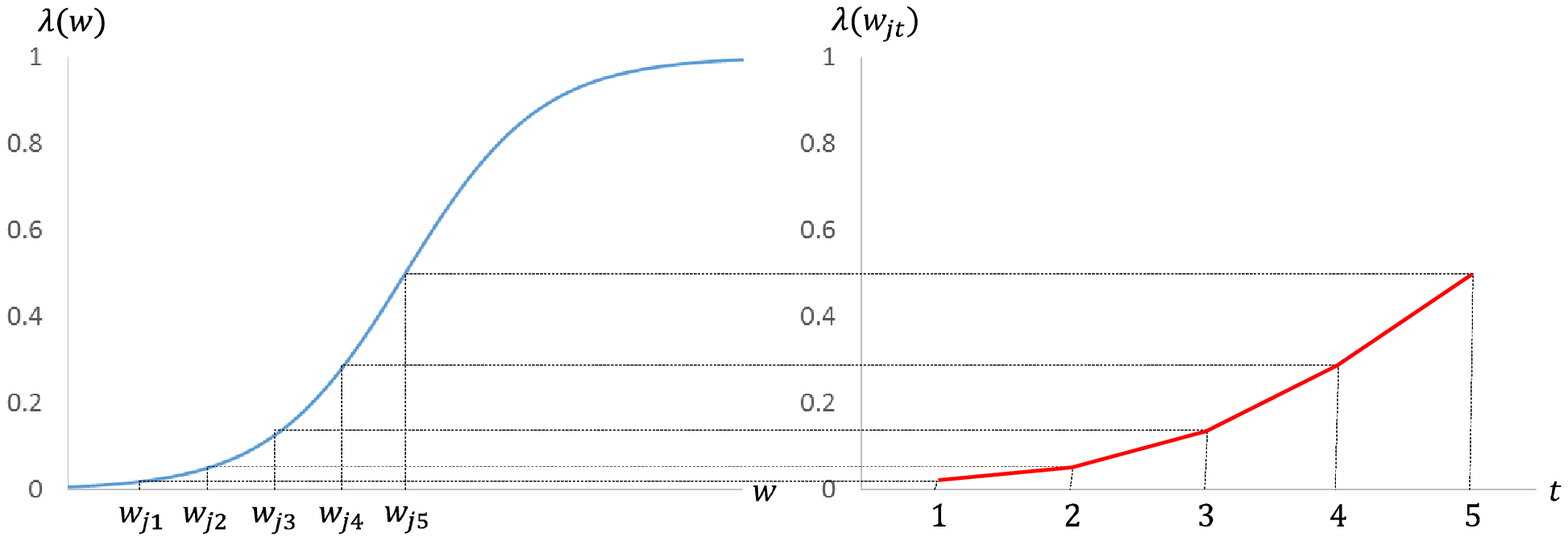} \\ (b) $(w_{j1},w_{j2},w_{j3},w_{j4},w_{j5}) = (-4,-3,-2,-1,0)$ \\\\
\includegraphics[scale=0.6]{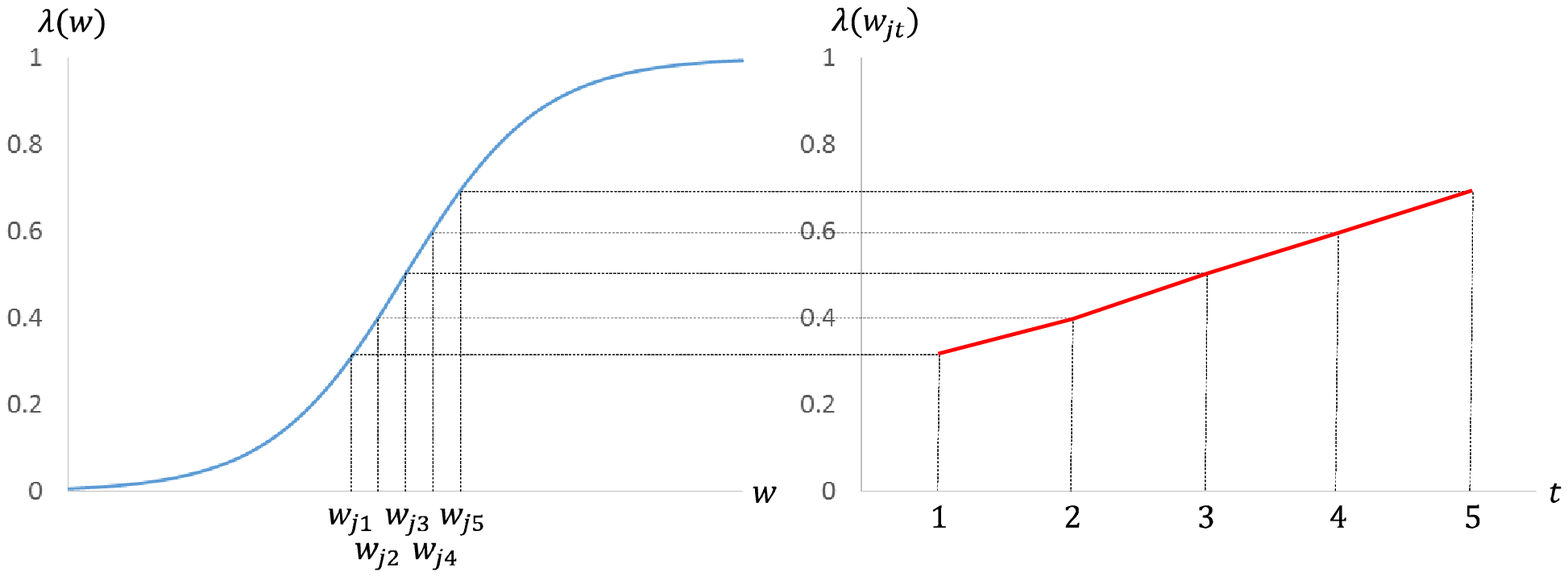} \\ (c) $(w_{j1},w_{j2},w_{j3},w_{j4},w_{j5}) = (-1,-0.5,0,0.5,1)$
\end{tabular}
\caption{Smoothness-constrained ICCs with $\gamma = 0$}
\label{fig:icc_ex}
\end{figure}

Fig.~\ref{fig:icc_ex} illustrates three examples of smoothness-constrained ICCs with $\gamma = 0$, where the figures on the left-hand side are graphs of the logistic function, and those on the right-hand side are the corresponding ICCs. It is clear that an S-shaped ICC can be created as shown in Fig.~\ref{fig:icc_ex}(a). In addition, although $w_{j1}, w_{j2}, \ldots, w_{j5}$ must be equally spaced points, the difficulty and discrimination of a test item can be adjusted. For instance, Fig.~\ref{fig:icc_ex}(b) and (c) correspond to difficult and undiscriminating items, respectively. These examples demonstrate that the smoothness constraints~\eqref{con:sc} keep each ICC close to a logistic curve with two parameters, namely difficulty and discrimination. Therefore, our smoothness constraints provide benefits to both PIRT and NIRT models in the sense that the shapes of nonparametric ICCs are restricted by means of parametric functions. 

\subsection{Formulation}
We begin by substituting $x_{jt} = \lambda(w_{jt})$ into the log-likelihood function~\eqref{eq:llf} as follows: 
\begin{align} 
\sum_{(i,j,t)  \in I \times J \times T} y_{it} \left( u_{ij} \log \lambda(w_{jt}) + (1 - u_{ij}) \log (1 - \lambda(w_{jt})) \right). \label{eq:llf2}
\end{align}
It then follows from Eq.~\eqref{eq:logf} that 
\begin{align}
    & u_{ij} \log \lambda(w_{jt}) + (1 - u_{ij}) \log (1 - \lambda(w_{jt})) \notag \\
=\, & u_{ij} \log \lambda(w_{jt}) + (1 - u_{ij}) \log \lambda(-w_{jt}) \notag \\
=\, & \log \lambda((2u_{ij} - 1)w_{jt}) \qquad (\because\,u_{ij} \in \{0,1\}) \notag \\
=\, & - \log(1 + \exp((1 - 2u_{ij})w_{jt})). \label{eq:trans}
\end{align}
Therefore, maximizing the log-likelihood function~\eqref{eq:llf2} is equivalent to minimizing 
\begin{align}
& \sum_{(i,j,t)  \in I \times J \times T} y_{it} \log(1 + \exp((1 - 2u_{ij})w_{jt})). 
\label{eq:logf2}
\end{align}
We show next that the smoothness constraints~\eqref{con:sc} can be converted into a set of linear constraints. 
\begin{proposition}
The smoothness constraints~(\ref{con:sc}) hold if and only if there exist $\bm{S} := (s_{jt})_{(j,t) \in J \times T}$ and $\bm{V} := (v_{jt})_{(j,t) \in J \times T}$ such that 
\begin{align}
& \quad \sum_{t \in T} (s_{jt} + v_{jt}) \le \gamma \quad (j \in J),  \label{con:sc2a} \\
& \quad s_{jt} - v_{jt} = w_{j,t+2} - 2w_{j,t+1} + w_{jt} \quad ((j,t) \in J \times T), \label{con:sc2b} \\
& \quad s_{jt} \ge 0,~v_{jt} \ge 0 \quad ((j,t) \in J \times T).  \label{con:sc2c} 
\end{align}
\label{prop1}
\end{proposition}
\noindent
\begin{proof}
Firstly, we suppose that the smoothness constraints~\eqref{con:sc} are satisfied by $\bar{\bm{W}} := (\bar{w}_{jt})_{(j,t) \in J \times T}$. 
We then set $\bar{\bm{S}} := (\bar{s}_{jt})_{(j,t) \in J \times T}$ and $\bar{\bm{V}} := (\bar{v}_{jt})_{(j,t) \in J \times T}$ as 
\[
\begin{cases}
\bar{s}_{jt} := \bar{w}_{j,t+2} - 2\bar{w}_{j,t+1} + \bar{w}_{jt},~\bar{v}_{jt} := 0 & \mbox{if}~\bar{w}_{j,t+2} - 2\bar{w}_{j,t+1} + \bar{w}_{jt} \ge 0, \\
\bar{s}_{jt} := 0,~\bar{v}_{jt} := -(\bar{w}_{j,t+2} - 2\bar{w}_{j,t+1} + \bar{w}_{jt}) & \mbox{otherwise}, \\
\end{cases}
\]
for each $(j,t) \in J \times T$. 
It then follows that constraints~\eqref{con:sc2a}--\eqref{con:sc2c} are satisfied by $\bar{\bm{S}}$, $\bar{\bm{V}}$, and $\bar{\bm{W}}$ because 
\[
\sum_{t \in T} (\bar{s}_{jt} + \bar{v}_{jt}) = \sum_{t \in T} |\bar{w}_{j,t+2} - 2\bar{w}_{j,t+1} + \bar{w}_{jt}| 
\le \gamma \quad (j \in J).
\]

Conversely, we suppose that constraints~\eqref{con:sc2a}--\eqref{con:sc2c} are satisfied by $\bar{\bm{S}}$, $\bar{\bm{V}}$, and $\bar{\bm{W}}$. 
The smoothness constraints~\eqref{con:sc} are then satisfied as follows: 
\[
\sum_{t \in T} |\bar{w}_{j,t+2} - 2\bar{w}_{j,t+1} + \bar{w}_{jt}| 
= \sum_{t \in T} |\bar{s}_{jt} - \bar{v}_{jt}|
\le \sum_{t \in T} (\bar{s}_{jt} + \bar{v}_{jt}) 
\le \gamma \quad (j \in J).
\]$\,$
\qed
\end{proof}

Consequently, our smoothness-constrained model minimizes Eq.~\eqref{eq:logf2} subject to constraints~\eqref{con:asg1}--\eqref{con:asg2}, \eqref{con:smh}, and \eqref{con:sc2a}--\eqref{con:sc2c}:
\begin{align}
\mathop{\mbox{minimize}}_{\scriptstyle \bm{S},\bm{V},\bm{W},\bm{Y}} & \quad \sum_{(i,j,t)  \in I \times J \times T} y_{it} \log(1 + \exp((1 - 2u_{ij})w_{jt})) \label{obj:SCM} \\
\mbox{subject~to} 
& \quad w_{jt} \le w_{j,t+1} \quad ((j,t) \in J \times T), \label{con1:SCM} \\
& \quad \sum_{t \in T} (s_{jt} + v_{jt}) \le \gamma \quad (j \in J),  \label{con2:SCM} \\
& \quad s_{jt} - v_{jt} = w_{j,t+2} - 2w_{j,t+1} + w_{jt} \quad ((j,t) \in J \times T), \label{con3:SCM} \\
& \quad s_{jt} \ge 0,~v_{jt} \ge 0 \quad ((j,t) \in J \times T), \label{con4:SCM} \\
& \quad \sum_{t \in T} y_{it} = 1 \quad (i \in I), \label{con5:SCM} \\
& \quad y_{it} \in \{0,1\} \quad ((i,t) \in I \times T). \label{con6:SCM} 
\end{align}

\subsection{EM algorithm}
We employ an EM algorithm to find a good-quality solution to the smoothness-constrained model~\eqref{obj:SCM}--\eqref{con6:SCM} efficiently. 
To this end, we introduce the decision variable $\bm{\pi} := (\pi_t)_{t \in T}$ for the sizes of ability classes, such that 
\begin{align} \label{con:size}
\sum_{t \in T} \pi_t = 1, \qquad \pi_t > 0 \quad (t \in T). 
\end{align}
The conditional probability of receiving response $\bm{u}_{i \cdot}$ from the $i$th examinee, given that s/he belongs to the $t$th ability class (i.e., $y_{it} = 1$), is expressed as 
\begin{align}
f(\bm{u}_{i \cdot} \mid \bm{w}_{\cdot t}) := \prod_{j \in J} \lambda(w_{jt})^{u_{ij}} (1 - \lambda(w_{jt}))^{1 - u_{ij}}. \notag
\end{align}
Accordingly, the marginal likelihood is calculated by a weighted sum of the form
\begin{align}
\sum_{t \in T} \pi_t f(\bm{u}_{i \cdot} \mid \bm{w}_{\cdot t}). \notag
\end{align}
The posterior probability of the $i$th examinee belonging to the $t$th ability class is given by Bayes' rule as follows: 
\begin{align}
\frac{\pi_t f(\bm{u}_{i \cdot} \mid \bm{w}_{\cdot t})}{\sum_{\tau \in T} \pi_{\tau} f(\bm{u}_{i \cdot} \mid \bm{w}_{\cdot \tau})}. 
\label{eq:pp}
\end{align}
Moreover, the complete-data log-likelihood function based on $\bm{Y}$ is formulated as follows:
\begin{align}
  & \log \left( \prod_{(i,t) \in I \times T} \left( \pi_t f(\bm{u}_{i \cdot} \mid \bm{w}_{\cdot t}) \right)^{y_{it}} \right) \notag \\
= & \underbrace{\sum_{(i,t) \in I \times T} y_{it} \log \pi_t}_{\rm (\ref{eq:cllf}a)} + \underbrace{\sum_{(i,t) \in I \times T} y_{it} \log f(\bm{u}_{i \cdot} \mid \bm{w}_{\cdot t})}_{\rm (\ref{eq:cllf}b)}. \label{eq:cllf} 
\end{align}

Our EM algorithm starts with some initial estimate of the ability classes 
\[
\bar{\bm{Y}} := (\bar{y}_{it})_{(i,t) \in I \times T}. 
\]
To obtain $\bar{\bm{Y}}$, one may use the number of test items that each examinee answered correctly. The EM algorithm then repeats the E-step (expectation step) and M-step (maximization step) to maximize the log-likelihood function~\eqref{eq:cllf}. 

The M-step substitutes $\bar{\bm{Y}}$ into the log-likelihood function~\eqref{eq:cllf} and then maximizes it. We firstly consider maximizing Eq.~(\ref{eq:cllf}a) subject to constraints~\eqref{con:size}. The method of Lagrange multipliers yields $\bar{\bm{\pi}} := (\bar{\pi}_t)_{t \in T}$ defined by 
\begin{align}
\bar{\pi}_t \leftarrow \frac{\sum_{i \in I} \bar{y}_{it}}{|I|}
\label{eq:Mstep2}
\end{align}
for each $t \in T$. 

Next, we focus on maximizing Eq.~(\ref{eq:cllf}b) after substituting Eq.~\eqref{eq:trans} into it. This is equivalent to solving the smoothness-constrained model~\eqref{obj:SCM}--\eqref{con6:SCM} with $\bm{Y} = \bar{\bm{Y}}$; that is, for each $j \in J$ we solve
\begin{align}
\mathop{\mbox{minimize}}_{\scriptstyle \bm{s}_{j \cdot},\bm{v}_{j \cdot},\bm{w}_{j \cdot}} & \quad \sum_{(i,t) \in I \times T} \bar{y}_{it} \log(1 + \exp((1 - 2u_{ij})w_{jt})) \label{obj:sSCM} \\
\mbox{subject~to} 
& \quad w_{jt} \le w_{j,t+1} \quad (t \in T), \label{con1:sSCM} \\
& \quad \sum_{t \in T} (s_{jt} + v_{jt}) \le \gamma, \label{con2:sSCM} \\
& \quad s_{jt} - v_{jt} = w_{j,t+2} - 2w_{j,t+1} + w_{jt} \quad (t \in T), \label{con3:sSCM} \\
& \quad s_{jt} \ge 0,~v_{jt} \ge 0 \quad (t \in T). \label{con4:sSCM} 
\end{align}
This problem minimizes a convex function subject to linear constraints, so it can be solved exactly and efficiently by standard nonlinear optimization software. The estimates obtained at this step are denoted by 
\[
\bar{\bm{W}} := (\bar{w}_{jt})_{(j,t) \in J \times T}. 
\]

The E-step updates $\bar{\bm{Y}}$ with its expected value based on the current estimates $\bar{\bm{W}}$ and $\bar{\bm{\pi}}$. 
This amounts to assigning the posterior probability~\eqref{eq:pp} as follows: 
\begin{align} \label{eq:Estep}
\bar{y}_{it} \leftarrow \frac{\bar{\pi}_t f(\bm{u}_{i \cdot} \mid \bar{\bm{w}}_{\cdot t})}{\sum_{\tau \in T} \bar{\pi}_{\tau} f(\bm{u}_{i \cdot} \mid \bar{\bm{w}}_{\cdot \tau})}
\end{align}
for all $(i,t) \in I \times T$. 
The E-step and M-step are repeated until a termination condition is satisfied. Our EM algorithm for solving the smoothness-constrained model~\eqref{obj:SCM}--\eqref{con6:SCM} is summarized as follows: 
\begin{description}
\item[Step 0]{\sffamily (Initialization)}~
Set $\bar{\bm{Y}}$ as an initial estimate, and go to Step 2. 
\item[Step 1]{\sffamily (E-Step)}~
Update $\bar{\bm{Y}}$ according to Eq.~\eqref{eq:Estep} for $(i,t) \in I \times T$. 
\item[Step 2]{\sffamily (M-Step)}~
Update $\bar{\bm{\pi}}$ according to Eq.~\eqref{eq:Mstep2} for $t \in T$.
Update $\bar{\bm{W}}$ by solving problem~\eqref{obj:sSCM}--\eqref{con4:sSCM} for $j \in J$. 
\item[Step 3]{\sffamily (Termination Condition)}~
Terminate the algorithm if a termination condition is satisfied. 
Otherwise, return to Step 1.
\end{description}

\section{Computational experiments}\label{sec:4}
The computational results reported in this section evaluate the effectiveness of our smoothness-constrained NIRT model.

\subsection{Experimental design}
\begin{figure}[tbh]
\begin{center}
\includegraphics[scale=0.4]{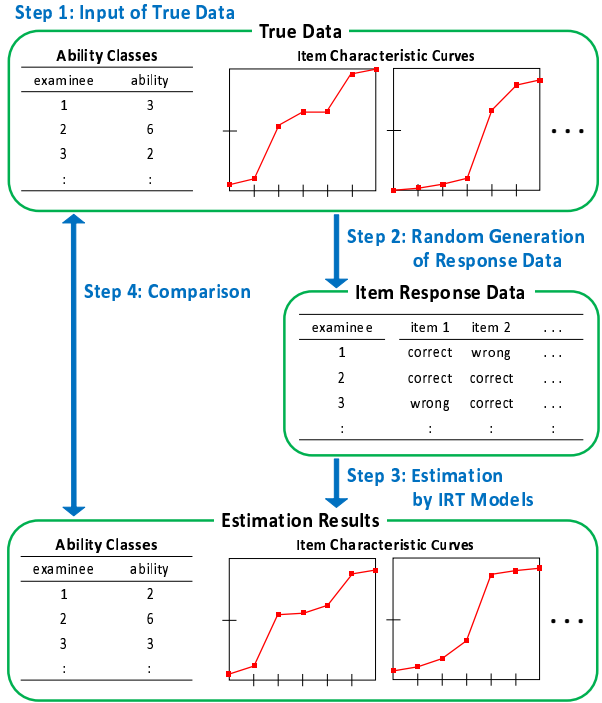}
\caption{Simulation process of model evaluation}
\label{Fig:Exper}
\end{center}
\end{figure}

We evaluated the estimation accuracy of IRT models through the simulation process illustrated in Fig.~\ref{Fig:Exper}. 

In Step 1, we randomly generate $\theta_i$ for $i \in I$ from a standard normal distribution. Next, we give a true ability class $t^*_i$ for $i \in I$ on the basis of $\theta_i$ and Table~\ref{tab:Rela}. For instance, if $-1.29 \le \theta_i < -0.81$, we set $t^*_i := 2$. The ranges of $\theta$ in Table~\ref{tab:Rela} were determined so that each ability class is assigned to approximately the same number of examinees. 

\begin{table}[htb]
\begin{center}
\caption{Relationship between ability class $t \in T$ and normal random number $\theta$}
\begin{tabular}{ccr} \hline\noalign{\smallskip}
$t$ & range of $\theta$ & median of $\theta$ \\ \noalign{\smallskip}\hline\noalign{\smallskip}
1 & $(-\infty, -1.29)$ & $-1.73$ \\
2 & $[-1.29, -0.81)$   & $-1.02$ \\
3 & $[-0.81, -0.49)$   & $-0.64$ \\
4 & $[-0.49, -0.23)$   & $-0.36$ \\
5 & $[-0.23, 0)$       & $-0.12$ \\
6 & $[0, 0.23)$        & $0.12$  \\
7 & $[0.23, 0.49)$     & $0.36$  \\
8 & $[0.49, 0.81)$     & $0.64$  \\
9 & $[0.81, 1.29)$     & $1.02$  \\
10 & $[1.29, \infty)$  & $1.73$  \\ \noalign{\smallskip}\hline
\end{tabular}
\label{tab:Rela}
\end{center}
\end{table}

We used two types of function to create the true ICCs of test items. One was the two-parameter logistic (2PL) model 
\begin{align}
p_j^{\rm 2PL}(\theta) := \frac{1}{1 + \exp(-1.7 a_j (\theta - b_j))},
\label{eq:2PL}
\end{align}
where $a_j$ and $b_j$ are parameters of discrimination and difficulty, respectively. These parameters were drawn from uniform distributions for which $a_j \in [0.5, 2.0]$ and $b_j \in [-1.5, 1.5]$. Similarly to \cite{No08}, the other type of function was the extended three-parameter normal ogive (3PN) model of order two:
\begin{align}
p_j^{\rm 3PN}(\theta) := \Phi(a_{j2} (\theta - b_j)^3 + \sqrt{3 a_{j1} a_{j2}} (\theta - b_j)^2 + a_{j1} (\theta - b_j)), 
\label{eq:3PN}
\end{align}
where $\Phi$ is the normal ogive, $a_{j1}$ and $a_{j2}$ are shape parameters, and $b_j$ is a difficulty parameter. These parameters were drawn from uniform distributions for which $a_{j1} \in [0.4, 0.8]$, $a_{j2} \in [0.1, 0.5]$, and $b_j \in [-0.5,0.5]$. The 3PN model is based on the assumption that examinees' abilities follow a bimodal distribution. Accordingly, the standard two-parameter logistic IRT models have difficulty in fitting ICCs defined by the 3PN model, whereas they can accurately fit ICCs defined by the 2PL model. 

When the true ICC of the $j$th item was defined by the 2PL model~\eqref{eq:2PL}, it was set as ${x}^*_{j1} := p_j^{\rm 2PL}(-1.73), {x}^*_{j2} := p_j^{\rm 2PL}(-1.02),\ldots, {x}^*_{j,10} := p_j^{\rm 2PL}(1.73)$ according to the median of $\theta$ in each range (see Table~\ref{tab:Rela}). The true ICCs defined by the 3PN model~\eqref{eq:3PN} were set in the same manner. The percentage of ICCs defined by the 3PN model is denoted by $\rho$, where $\rho \in \{0\%, 20\%, 50\%\}$ similarly to \cite{No08}. 
For instance, if $|J| = 60$ and $\rho = 20\%$, true ICCs of 12 items are created by the 3PN model. 

In Step 2, binary item-response data $\bm{U}$ are generated randomly based on the true ability classes and ICCs specified at Step 1. Specifically, examinees in the $t$th ability class answered the $j$th item correctly with probability ${x}^*_{jt}$. 

In Step 3, ability classes and ICCs are estimated from the item-response data $\bm{U}$ by using the following IRT models: 
\begin{description}
\item[2PLM:] two-parameter logistic model,
\item[MHM:] monotone-homogeneity model~\eqref{obj:MHM}--\eqref{con4:MHM},
\item[SCM($\gamma$):] smoothness-constrained model~\eqref{obj:SCM}--\eqref{con6:SCM} with $\gamma \in \{0,1,2,4\}$. 
\end{description}
We used the {\sf irtoys} package in R 3.1.2~(\url{http://www.R-project.org}) to perform the 2PLM computations. Here, the continuous-valued ability $\theta$ estimated by 2PLM was converted into an ability class $t$ according to Table~\ref{tab:Rela} for comparison purposes. To solve MHM and SCM, we implemented the EM algorithm by using MATLAB~R2015a~(\url{https://www.mathworks.com/products/matlab.html}), in which problem~\eqref{obj:sSCM}--\eqref{con4:sSCM} was solved by the {\sf fmincon} function in the MATLAB optimization toolbox. The initial estimate $\bar{\bm{Y}}$ was set by dividing examinees equally into 10 ability classes according to the number of correct answers. Every time $\bar{\bm{Y}}$ was updated in the EM algorithm, the ability classes of examinees were determined temporarily as $\hat{t}_i := \arg \max\{\bar{y}_{it} \mid t \in T\}$ for $i \in I$.  The algorithm was terminated if $\hat{t}_i$ remained the same as the previous one for all $i \in I$. 

In Step 4, we evaluate the estimation accuracy of the IRT models by comparing the true data (Step 1) with the estimates (Step 3). Specifically, the root-mean-square error (RMSE) of the ability classes is calculated as 
\begin{align} \notag
\sqrt{\frac{\sum_{i \in I} (t^*_i - \hat{t}_i)^2}{|I|}},
\end{align}
where $\hat{t}_i$ is the estimated ability class. 
The RMSE of the ICCs is calculated as 
\begin{align} \notag
\sqrt{\frac{\sum_{(j,t) \in J \times T} (x^*_{jt} - \hat{x}_{jt})^2}{|J||T|}}, 
\end{align}
where $\hat{x}_{jt}$ is the estimated probability of a correct answer. 
We repeated these steps 10 times and show the average RMSEs in the following section. 

\subsection{Computational results}
Tables~\ref{tab:icc}~and~\ref{tab:abl} give the RMSEs of the ability classes and ICCs for the 12 experimental conditions. Here, the number of examinees was $|I| \in \{1000, 3000\}$, and the number of test items was $|J| \in \{30, 60\}$. Because the ordinal scale of neural test theory grades examinees into roughly 10 classes~(\citealt{Sh07,Sh08}), the number of ability classes was $|T| = 10$. Note that the minimum RMSEs for each experimental condition are given in bold face in the tables.

We firstly focus on the accuracy of ICC estimation (Table~\ref{tab:icc}). It is reasonable that 2PLM estimated ICCs very accurately when the percentage of 3PN ICCs was $\rho = 0\%$. However, the estimation accuracy of 2PLM was greatly reduced by increasing the percentage of 3PN ICCs. Indeed, when $\rho \ge 20\%$, the RMSEs were often smaller for MHM than for 2PLM. We also note that the estimation accuracy of MHM increased with the amount of item-response data. For instance, the RMSEs of MHM for $(|I|,|J|) = (1000,30)$ were at least 0.057, and those for $(|I|,|J|) = (3000,60)$ were 0.034.

The largest RMSEs of the ICCs were those for SCM(0) in almost all cases because the shapes of the ICCs were tightly restricted by the smoothness constraints with $\gamma = 0$. In contrast, SCM(1) frequently obtained higher estimation accuracy than did MHM, especially for $\rho \le 20\%$. The smallest RMSEs were attained by SCM(2) in many cases, whereas only for $(|J|,\rho) = (60,50\%)$ were those provided by SCM(4).

We next move on to the accuracy of ability-class estimation (Table~\ref{tab:abl}). As in Table~\ref{tab:icc}, the estimation accuracy of 2PLM reduced markedly with the percentage of 3PN ICCs. In contrast, MHM made relatively accurate estimates when $(|J|,\rho) = (60,50\%)$. Very large RMSEs were still provided by SCM(0) as in Table~\ref{tab:icc}. Meanwhile, the smallest RMSEs were often obtained by SCM(1) with $|J| = 30$, and those were obtained by SCM(2) with $|J| = 60$. 

The results in Tables~\ref{tab:icc}~and~\ref{tab:abl} confirm that smoothness constraints are very effective in improving the estimation accuracy of NIRT models. In particular, we may say that SCM(2) delivers the best estimation performance on the whole. 

\begin{table}
\begin{center}
\caption{Root-mean-square error of item-characteristic curve estimation}
\begin{tabular}{rrrrrrrrr} \hline\noalign{\smallskip}
$|I|$ & $|J|$ & $\rho$ & 2PLM & MHM & SCM(0) & SCM(1) & SCM(2) & SCM(4) \\ \noalign{\smallskip}\hline\noalign{\smallskip}
1000 & 30  & 0\%    &{\bf 0.031} & 0.057 & 0.054 & 0.036 & 0.038 & 0.046       \\
     &     & 20\%   &0.049       & 0.057 & 0.073 & 0.042 & {\bf 0.040} & 0.046       \\
     &     & 50\%   &0.075       & 0.058 & 0.102 & 0.052 & {\bf 0.044} & 0.048       \\ \noalign{\smallskip}
     & 60  & 0\%    &{\bf 0.030} & 0.049 & 0.066 & 0.037 & 0.034 & 0.041       \\
     &     & 20\%   &0.053       & 0.047 & 0.078 & 0.043 & {\bf 0.036} & 0.040       \\
     &     & 50\%   &0.082       & 0.050 & 0.115 & 0.059 & {\bf 0.044} & {\bf 0.044} \\ \noalign{\smallskip}
3000 & 30  & 0\%    &{\bf 0.017} & 0.041 & 0.056 & 0.025 & 0.026 & 0.034       \\
     &     & 20\%   &0.046       & 0.041 & 0.069 & 0.032 & {\bf 0.028} & 0.034       \\
     &     & 50\%   &0.073       & 0.041 & 0.102 & 0.044 & {\bf 0.033} & 0.034       \\ \noalign{\smallskip}
     & 60  & 0\%    &0.031       & 0.034 & 0.064 & 0.027 & {\bf 0.023} & 0.028       \\
     &     & 20\%   &0.056       & 0.034 & 0.080 & 0.038 & {\bf 0.027} & 0.030       \\
     &     & 50\%   &0.083       & 0.034 & 0.116 & 0.058 & 0.033 & {\bf 0.029}       \\ \noalign{\smallskip} \hline
\end{tabular}
\label{tab:icc}
\end{center}
\end{table}
\begin{table}
\begin{center}
\caption{Root-mean-square error of ability-class estimation}
\begin{tabular}{rrrrrrrrr} \hline\noalign{\smallskip}
$|I|$ & $|J|$ & $\rho$ & 2PLM & MHM & SCM(0) & SCM(1) & SCM(2) & SCM(4) \\ \noalign{\smallskip}\hline\noalign{\smallskip}
1000 & 30  & 0\%    & 0.814 & 0.884 & 0.814 & {\bf 0.785} & 0.801 & 0.845 \\
     &     & 20\%   & 0.838 & 0.886 & 0.890 & {\bf 0.799} & 0.816 & 0.849       \\
     &     & 50\%   & 0.956 & 0.959 & 1.137 & {\bf 0.883} & 0.890 & 0.922       \\ \noalign{\smallskip}              
     & 60  & 0\%    & 0.629 & 0.680 & 0.747 & 0.611 & {\bf 0.606} & 0.642       \\
     &     & 20\%   & 0.704 & 0.680 & 0.812 & 0.642 & {\bf 0.624} & 0.648       \\
     &     & 50\%   & 0.914 & 0.715 & 1.176 & 0.740 & {\bf 0.676} & 0.700       \\ \noalign{\smallskip}              
3000 & 30  & 0\%    & 0.808 & 0.877 & 0.841 & {\bf 0.787} & 0.803 & 0.850       \\
     &     & 20\%   & 0.831 & 0.872 & 0.881 & {\bf 0.792} & 0.801 & 0.841       \\
     &     & 50\%   & 0.982 & 0.929 & 1.150 & 0.871 & {\bf 0.869} & 0.905       \\ \noalign{\smallskip}              
     & 60  & 0\%    & 0.642 & 0.629 & 0.739 & 0.585 & {\bf 0.576} & 0.603       \\
     &     & 20\%   & 0.747 & 0.642 & 0.828 & 0.638 & {\bf 0.599} & 0.621       \\
     &     & 50\%   & 0.997 & 0.694 & 1.190 & 0.761 & {\bf 0.666} & 0.675       \\ \noalign{\smallskip} \hline
\end{tabular}
\label{tab:abl}
\end{center}
\end{table}

Table~\ref{tab:time} gives the computation times (in seconds) required for estimating the IRT models. The computations of 2PLM were very rapid mainly because each ICC involved only two parameters. The computations of SCM($\gamma$) became slower as $\gamma$ or $\rho$ increased. The computation time for estimating MHM was always the longest among all the models. These results suggest that when many 3PN ICCs are estimated subject to loose smoothness constraints, our EM algorithm takes a relatively long time to terminate. 

\begin{table}
\begin{center}
\caption{Computation times (s)}
\begin{tabular}{rrrrrrrrr} \hline\noalign{\smallskip}
$|I|$ & $|J|$ & $\rho$ & 2PLM & MHM & SCM(0) & SCM(1) & SCM(2) & SCM(4) \\ \noalign{\smallskip}\hline\noalign{\smallskip}
1000 & 30  & 0\%    &3.3  & 172.2 & 42.6 & 43.3 & 75.6 & 140.2       \\
     &     & 20\%   &3.2  & 173.0 & 56.7 & 62.5 & 83.7 & 129.4       \\
     &     & 50\%   &3.1  & 200.8 & 77.0 & 106.3 & 115.6 & 158.2     \\ \noalign{\smallskip}
     & 60  & 0\%    &8.2  & 383.1 & 81.8 & 113.2 & 118.6 & 186.5     \\
     &     & 20\%   &8.2  & 317.4 & 144.0 & 166.7 & 149.0 & 214.9    \\
     &     & 50\%   &8.4  & 409.6 & 224.6 & 265.2 & 220.0 & 259.9    \\ \noalign{\smallskip}
3000 & 30  & 0\%    &13.8 & 1140.0& 88.3 & 128.3 & 310.2 & 597.9    \\
     &     & 20\%   &14.2 & 1026.5& 171.4 & 269.5 & 326.5 & 542.3   \\
     &     & 50\%   &13.7 & 1273.7& 181.6 & 281.2 & 323.0 & 672.6   \\ \noalign{\smallskip}
     & 60  & 0\%    &36.7 & 1743.7& 214.0 & 355.2 & 364.8 & 830.3   \\
     &     & 20\%   &36.0 & 1949.8& 443.8 & 765.2 & 555.2 & 825.2   \\
     &     & 50\%   &35.8 & 3053.2& 487.5 & 1008.0 & 898.0 & 1034.0 \\ \noalign{\smallskip} \hline
\end{tabular}
\label{tab:time}
\end{center}
\end{table}

\begin{figure}
\centering
\begin{tabular}{ccc}
\includegraphics[scale=0.4]{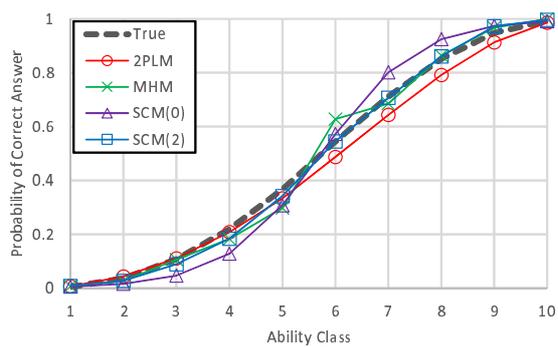} \\ (a) test item $j = 7$ \\\\
\includegraphics[scale=0.4]{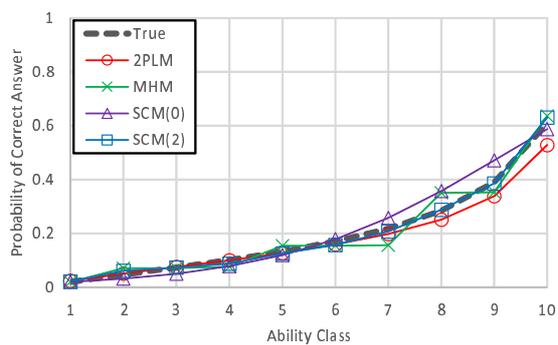} \\ (b) test item $j = 9$ \\\\
\includegraphics[scale=0.4]{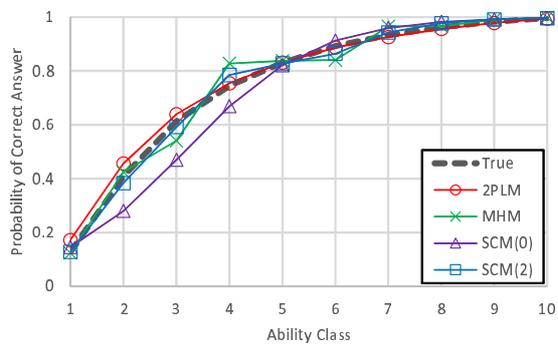} \\ (c) test item $j = 27$
\end{tabular}
\caption{Estimated ICCs and true 2PL ICCs ($(|I|,|J|,\rho) = (3000,60,0\%))$}
\label{fig:icc2pl}
\end{figure}

\begin{figure}
\centering
\begin{tabular}{ccc}
\includegraphics[scale=0.4]{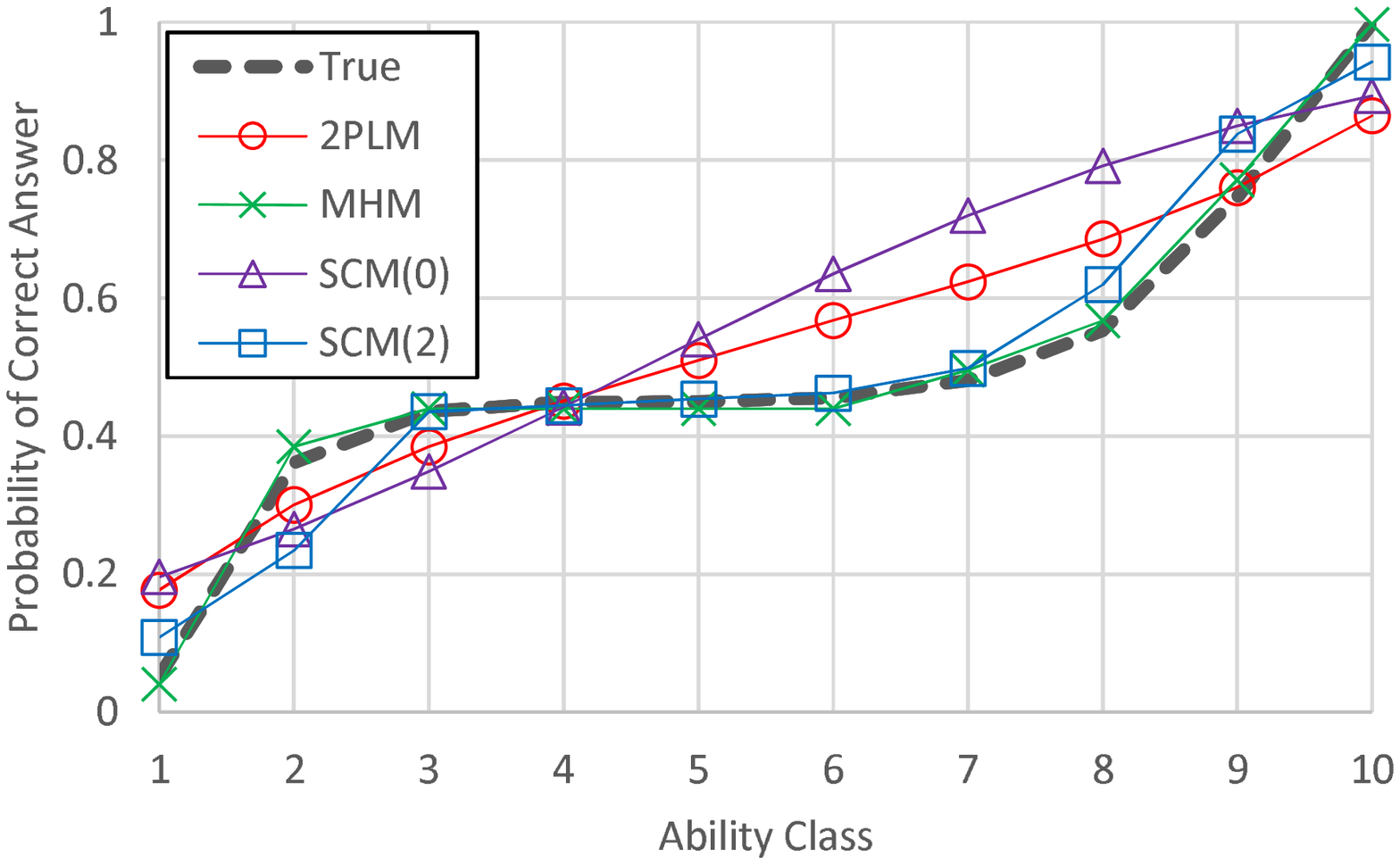} \\ (a) test item $j = 33$ \\\\
\includegraphics[scale=0.4]{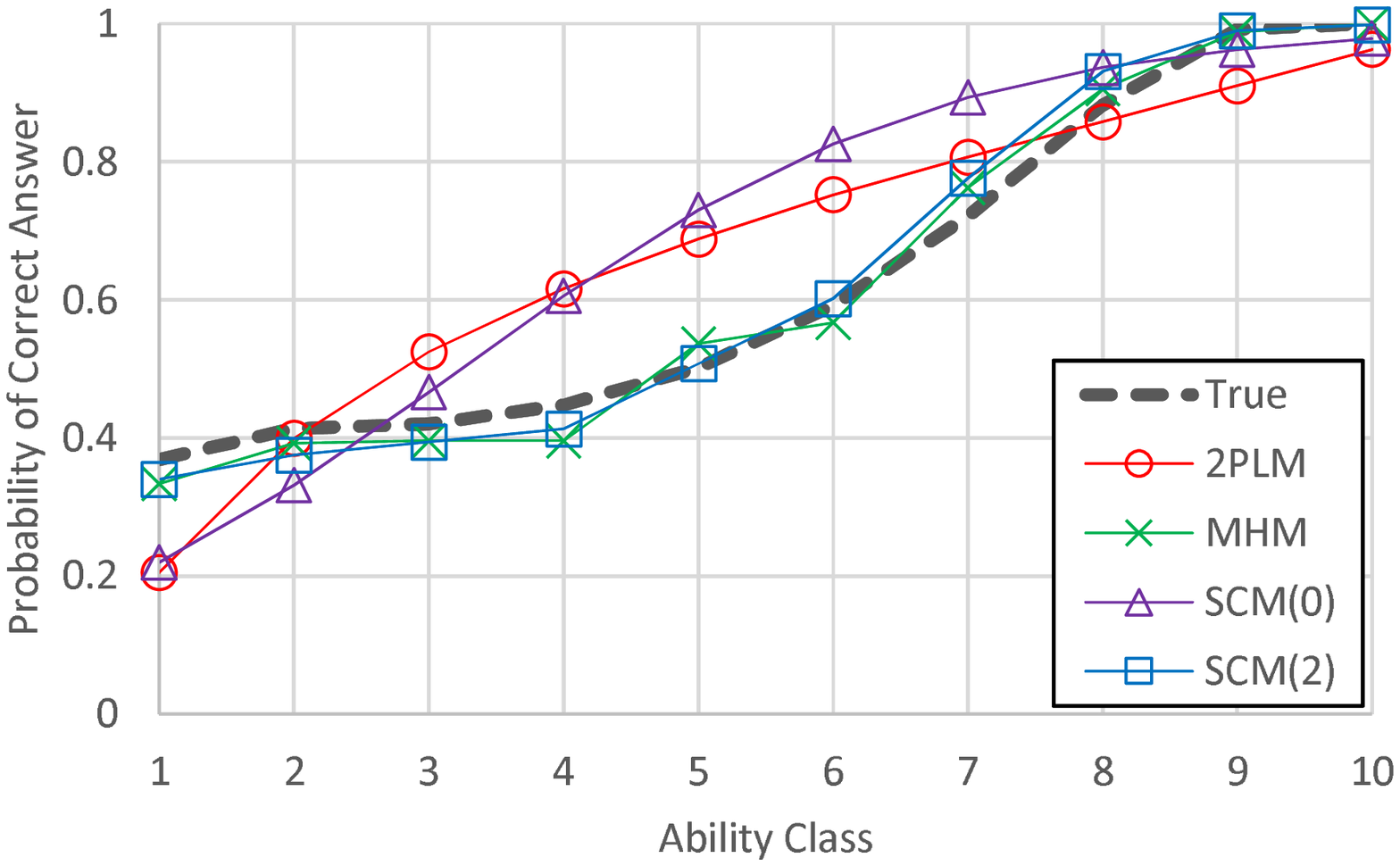} \\ (b) test item $j = 36$ \\\\
\includegraphics[scale=0.4]{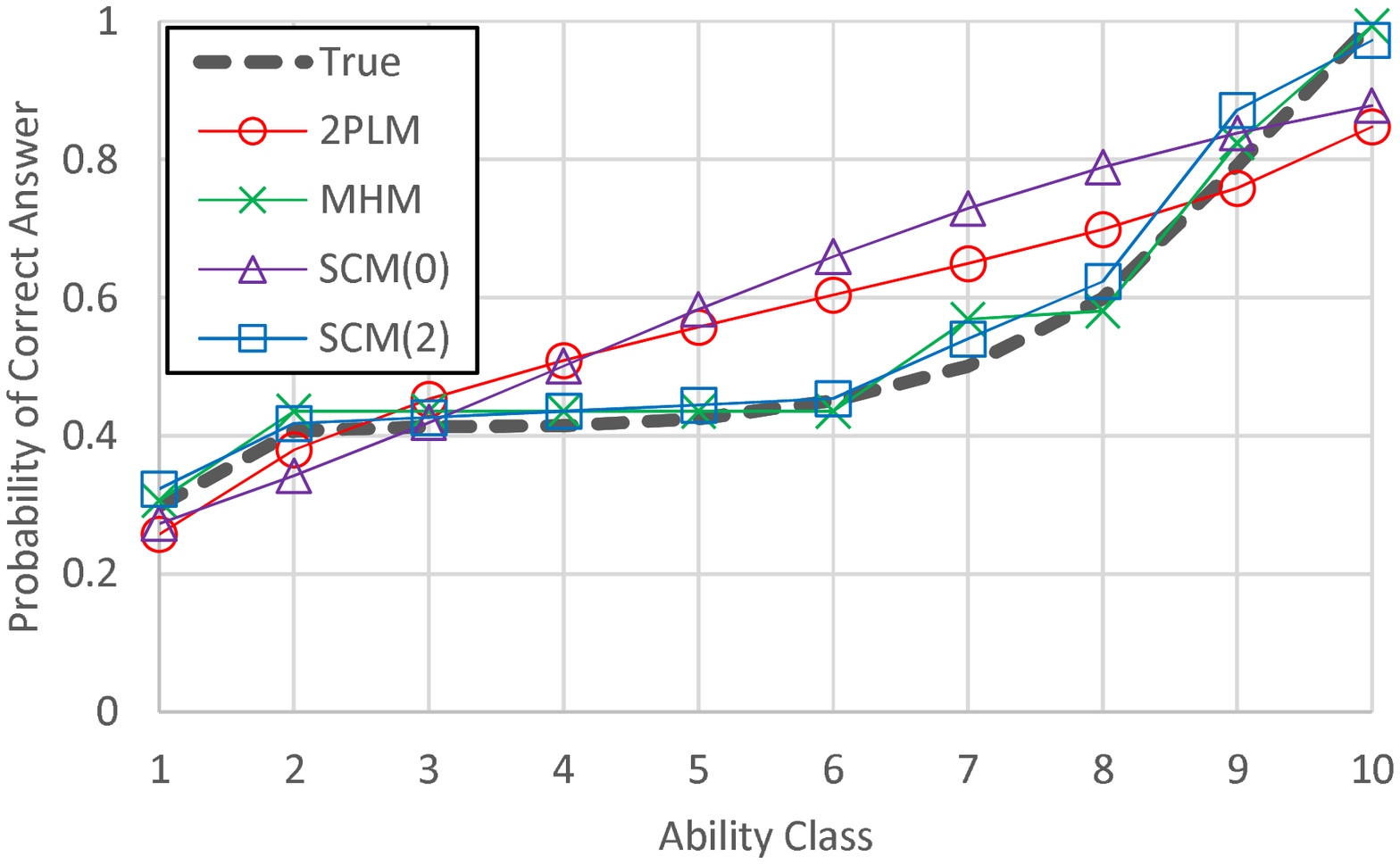} \\ (c) test item $j = 59$
\end{tabular}
\caption{Estimated ICCs and true 3PN ICCs ($(|I|,|J|,\rho) = (3000,60,50\%))$}
\label{fig:icc3pn}
\end{figure}

Figs.~\ref{fig:icc2pl}~and~\ref{fig:icc3pn} show illustrative examples of estimated ICCs together with the true ICCs. We firstly focus on Fig.~\ref{fig:icc2pl}, where $(|I|,|J|,\rho) = (3000,60,0\%)$, and the true ICCs were defined by the 2PL model. 
As expected, 2PLM fitted the true ICCs well. The MHM also fitted the true ICCs well, but it is noteworthy that the ICCs estimated by MHM moved around the true ICCs: for example, for the ability classes $t = 7, 8$, and 9 in Fig.~\ref{fig:icc2pl}(b) and $t= 3, 4$, and 6 in Fig.~\ref{fig:icc2pl}(c). The SCM(0) deviated partly from the true ICCs: for example, for the ability classes $t= 3, 4, 7$, and 8 in Fig.~\ref{fig:icc2pl}(a). In contrast, the true ICCs and those estimated by SCM(2) were almost the same because the nonparametric ICCs were made moderately less flexible by the smoothness constraints with $\gamma = 2$. 

We next move on to Fig.~\ref{fig:icc3pn}, where $(|I|,|J|,\rho) = (3000,60,50\%)$ and the true ICCs were defined by the 3PN model. It is clear that neither 2PLM nor SCM(0) fitted the true ICC because these models create only logistic curves. In contrast, MHM and SCM(2) estimated the shapes of the true ICCs accurately, except that SCM(2) underestimated the probability of a correct answer for the ability class $t = 2$ in Fig.~\ref{fig:icc3pn}(a). 


\section{Conclusions}\label{sec:5}
We devised an NIRT model with monotone homogeneity and smoothness constraints based on ordered latent classes. Our smoothness constraints avoid overfitting of nonparametric ICCs by smoothing them based on logistic curves. We also developed an EM algorithm for our smoothness-constrained NIRT model to estimate the nonparametric ICCs and the latent abilities of examinees efficiently. 

The computational results demonstrated the effectiveness of our model in comparison to the two-parameter logistic model and the monotone-homogeneity model. Indeed, our model obtained more accurate estimation results than did the two-parameter logistic model when the latent abilities of examinees for some test items followed bimodal distributions. Moreover, our model outperformed the monotone-homogeneity model because of the effect of the smoothness constraints. 

The contributions of this research are twofold. Firstly, we formulated the smoothness-constrained NIRT model as a mathematical optimization problem. Although we implemented the EM algorithm for the purpose of efficient computation, high-performance mixed-integer optimization algorithms could be used to compute an optimal solution to the problem~(\citealt{Bi12}). Secondly, we validated the utility of shape restrictions on nonparametric ICCs in avoiding an unstable ICC estimation specific to NIRT models. A future direction of study will be to impose other shape restrictions on nonparametric ICCs and evaluate their effectiveness. 

Although PIRT models are used commonly by many testing companies, it is known that they do not always fit the actual item-response data well. In contrast, NIRT models have the potential to estimate various forms of ICCs, and the estimation performance could be improved by incorporating the smoothness constraints. So, we expect that our research will extend the usefulness of NIRT models. 

\begin{acknowledgements}
This work was supported by JSPS KAKENHI Grant Number 26750114.
\end{acknowledgements}



\end{document}